# Triple-pulse X-ray spot size measurement of the Dragon-II linear induction accelerator

Yi Wang[*], Qin Li, Heng-Song Ding, Jin-Ming Cheng, Yun-Long Liu, Jin-Shui Shi, Jian-Jun Deng

Key Laboratory of Pulsed Power, Institute of Fluid Physics, China Academy of Engineering Physics, Mianyang 621900, Sichuan Province, China

[*]Corresponding author: wangyi_caep@163.com

**Abstract:** The Dragon-II linear induction accelerator is able to provide triple-pulse electron beams with an adjustable pulse spacing at the minimum of 300 ns. As a main determinant of the image quality, the X-ray spot size is usually quoted as an evaluation of the resolving power. The pinhole imaging method is applied to measure the two-dimensional spatial distribution of the triple-pulse X-ray source, by which the azimuthal asymmetry of the source can be analyzed. The spot size of each pulse is characterized by both the full-width at half-maximum and the LANL definition. The weighted RMS radius at different sum ratios of photon intensity is also calculated, which directly takes into account the contribution of every single photon within the boundary of the spot.

## 1. Introduction

In order to investigate the hydrodynamic process of high explosives, electron beam pulses are generated and accelerated to ~MeV energy in the linear induction accelerator (LIA) and finally focused onto a high-Z convertor to produce intense X-ray photons through the bremsstrahlung radiation [1-5]. The Dragon-II LIA is able to provide triple-pulse electron beams at a MHz repetition rate. The electron beam of each pulse has the energy of 18~20 MeV, the current of ~2 kA and the pulse width of ~70 ns [6]. The pulse spacing is adjustable and at a minimum of 300 ns. The spot size of the X-ray source is usually quoted as an evaluation of the resolving ability for the LIA. The focal size of the source spot is not only strongly dependent on the electron beam parameters, such as energy-spread and emittance [7], but also closely related to the hydrodynamic process of beam-target interaction [8]. A good knowledge of the source spot is of great importance not only for resolving the object, but also for the optimization of the converter target.

Due to a strong penetrating through materials for X-ray photons especially those with the energy of ~ MeV, it is not an easy task to precisely measure the focal size of the radiographic source spot. Different techniques have been proposed for high-energy X-ray spot size measurement, by which different information of the source spot was obtained [9-11]. The pinhole imaging method provides a full two-dimensional spatial distribution. The image obtained by the slit method actually denotes a projection of the spot orthogonal to the slit, i.e. the LSF. The edge imaging method uses an opaque shield to block part of source photons, by which the penumbral image obtained provides the edge-spread function (ESF) of the source. In order to ease alignment, a roll-bar that has a large radius of the roll curving edge is usually employed to replace the knife-edge shield. The ESF is actually the convolution of the LSF with a step function centered at the edge, so the differentiation of the ESF yields the LSF along the direction perpendicular to the edge.

The source spot of the bremsstrahlung radiation is generally azimuth-asymmetric because of corkscrew oscillation [12] of the electron beam during transporting and focusing, inaccurate alignments of the solenoid field and non-uniform of the converter target. Comparing with the slit and edge techniques, the pinhole imaging method is preferred for the spot size measurement since it has the advantage of resolving azimuthal asymmetry of the source. In this paper, the pinhole imaging method is applied to diagnostic the triple-pulse X-ray source spot produced by the Dragon-II LIA.

## 2. Principle

In the high-energy flash-radiography experiment, photons of the bremsstrahlung radiation from electrons striking a heavy-metal target penetrate through the object placed in the light path, the spatial distribution of which is recorded by the image-receiving system. This process of projection radiography can be described as

$$i(x, y) = s(x, y) * o(x, y) * r(x, y), \tag{1}$$

where $i(x, y)$ is the recorded image; $s(x, y)$ is the point-spread function (PSF) of the light source; $o(x, y)$ represents the point-projection imaging of the object; $r(x, y)$ denotes the blur of the image-receiving system, which corresponds to the response of the detector to a point source; the sign $*$ denotes the convolution operation.

The simple product relation of the modulation transfer function (MTF) can be obtained by making a Fourier transform of Eq. (1), i.e.

$$I(f_x, f_y) = S(f_x, f_y) \cdot O(f_x, f_y) \cdot R(f_x, f_y), \tag{2}$$

where I, S, O, R represent MTFs of each term in Eq. (1); $f_x$ and $f_y$ are the spatial frequencies. The MTF is defined as the modulus of the Fourier transform of the PSF

$$\text{MTF}(f_x, f_y) = \left| \int_{-\infty}^{+\infty} \int_{-\infty}^{+\infty} \text{PSF}(x, y) \exp\left[-i2\pi\left(f_x x + f_y y\right)\right] \mathrm{d}x \mathrm{d}y \right|, \tag{3}$$

which is a two-dimensional space of the spatial frequency. The slice of the MTF along a certain direction (take the $f_x$-direction for example, i.e. $f_y = 0$) is given by

$$\text{MTF}(f_x, 0) = \left| \int_{-\infty}^{+\infty} \mathrm{d}x \int_{-\infty}^{+\infty} \text{PSF}(x, y) \exp\left(-i2\pi f_x x\right) \mathrm{d}y \right|$$

$$= \left| \int_{-\infty}^{+\infty} \text{LSF}(x) \exp\left(-i2\pi f_x x\right) \mathrm{d}x \right|, \tag{4}$$

which yields the Fourier transform of the line-spread function (LSF) in the orthogonal projection.

The simplest metric to characterize the spot size is to use full-width at half-maximum (FWHM). Apparently, it does not distinguish different distributions of the source spot. Instead of considering the width at half peak value of spatial distribution, the spot size in the LANL definition ($D_L$) takes into consideration a particular level of the MTF in the spatial frequency space [13]. The MTF analyzes each component of the image as a low-pass filter for spatial information. Practically, $D_L$ is defined as the diameter of an equivalent uniform disk that has the same spatial frequency at 50% of the MTF peak value ($f_{50\% \text{MTF}}$) as the source spot:

$$D_L = \frac{0.705}{f_{50\% \text{MTF}}}. \tag{5}$$

The relation between PSF, LSF and MTF varies depending on the spot distribution. Fig. 1

compares the PSFs, the LSFs and the MTFs for typical distributions which are commonly referred to, including uniform disk (KV), Gaussian (GS), Bennett (BNT) and Quasi-Bennett (QBNT) models [14]. The curves in the figure have the same PSF FWHM of 1 mm. It shows a trend that a more centralized PSF has a narrower FWHM for the LSF and a larger $f_{50\% \, \text{MTF}}$ for the MTF, which denotes a smaller $D_L$. The ratios of LSF FWHM to PSF FWHM are 0.866 for KV, 1 for GS, 1.19 for BNT and 1.31 for QBNT. According to the LANL definition, the spot size $D_L$ is equal to the PSF FWHM for KV, which is calculated to be 1.60 times its PSF FWHM for GS, 2.70 times for BNT and 4.14 times for QBNT.

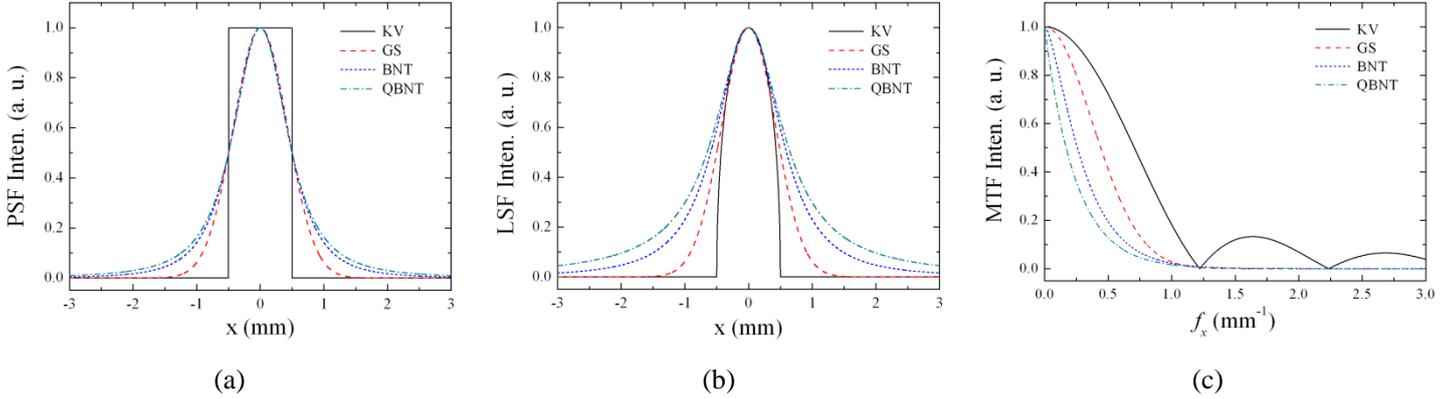

(a)            (b)            (c)

Fig. 1 (Color online) Comparison of PSFs, LSFs and MTFs for various distributions with PSF FWHM of 1 mm. (a) PSF curves; (b) LSF curves; (c) MTF curves.

## 3. Pinhole imaging method and experimental setup

Theoretically, the aperture of the pinhole and the width of the slit ought to be as small as possible for a clear imaging. The curvature radius of the roll-bar should be large enough to construct a quasi-knife-edge. An ideal object for imaging is defined to be infinitesimally thin and completely opaque, which cannot be realized for X-ray photons. The object needs to be thick enough to make an "opaque" sheet. And the pinhole aperture or the slit width cannot be too small in order to provide a view field covering enough area of the source. All these issues will induce errors in the X-ray spot size measurement.

A leading collimator is placed just in front of the radiographic source as a radiation shield of the area away from the central field. A 65-mm-thick tungsten bar with a pinhole of 0.47 mm diameter is precisely placed along the central axis. The image receiving system consists of an LYSO scintillator screen, a flat mirror titled at 45 deg, and a framing camera to record each pulse image. The camera is composed of an optical framing element and several intensifier CCD cameras. The gating time of the camera is controlled by a high-speed micro-channel plank. Precisions of the trigger time and the exposure time are less than 5 ns. By inputting external trigger synchronization signals, the camera records each pulse image separately.

The source-pinhole distance is $L_1$=1119 mm, and the screen-pinhole distance is $L_2$=4681 mm. Hence the geometrical magnification is calculated to be M=$L_2$/$L_1$=4.183. The pinhole block is placed much closer to the source to obtain a large magnification so as to acquire image with a better resolution and reduce the screen blur effect. According to the parameters of the pinhole object and the experimental alignment, the view field width for the source reaches 15.7 mm, which is about ten times the source FWHM, so the collimating effect of the pinhole block should be quite limited. In the experiment, the spot size measurements are conducted with the electron beam energy of ~ 18.9 MeV and current of ~ 2.05 kA, trying to maintain a steady and identical state for all pulses. In order to correct pixel-to-pixel variations of the screen sensitivity and dark noise of the camera, a standard procedure for gain and off modifications is applied for the image processing [15].

## 4. Results and discussions

Figure 2 shows typical images of the source spot obtained by the pinhole imaging method (No. 5295). Both the boundaries of half peak value and 90% photon intensity sum (PIS) are outlined. The FWHM of the spot image is given by the diameter of an equivalent disk that has the same area within the half-peak contour. Then the PSF FWHM of the source spot is obtained by dividing the geometrical magnification into the image FWHM, which is 1.33 mm for Pulse A, 1.41 mm for Pulse B and 1.50 mm for Pulse C. As shown in Fig. 3, each spot image is projected into the x and y directions, yielding two LSF curves perpendicular to the projections.

Table 1 shows the results of spot size measurements, which contains not only FWHM of the PSF but also FWHMs of two LSFs for each spot. The difference between FWHMs of LSF(x) and LSF(y) reveals the spot asymmetry. Apparently, the spot symmetry of Pulse A is better than those of Pulse B and Pulse C on the whole. The FWHM ratio of LSF to PSF is also calculated for each pulse, which, on average, is 1.34 for Pulse A, 1.27 for Pulse B and 1.31 for Pulse C. It reveals that the spatial distribution of the spot is close to QBNT.

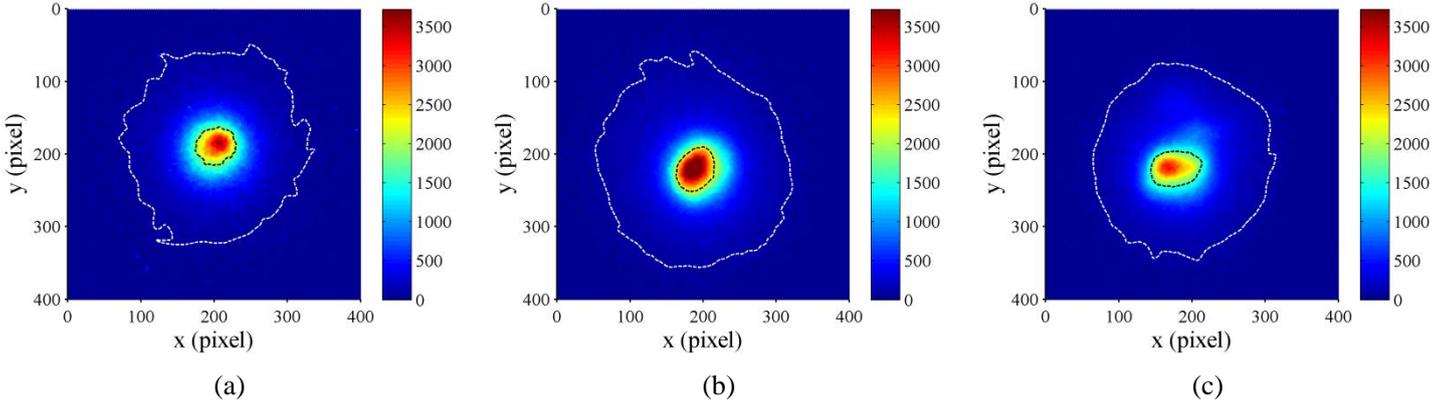

(a)                                      (b)                                      (c)

Fig. 2 (Color online) Typical images of triple-pulse X-ray spot by the pinhole method (No. 5295).
(a) Pulse A; (b) Pulse B; (c) Pulse C. The inside curve (in black dash) and the outside curve (in white dash) denote the boundaries of half peak and 90%PIS, respectively.

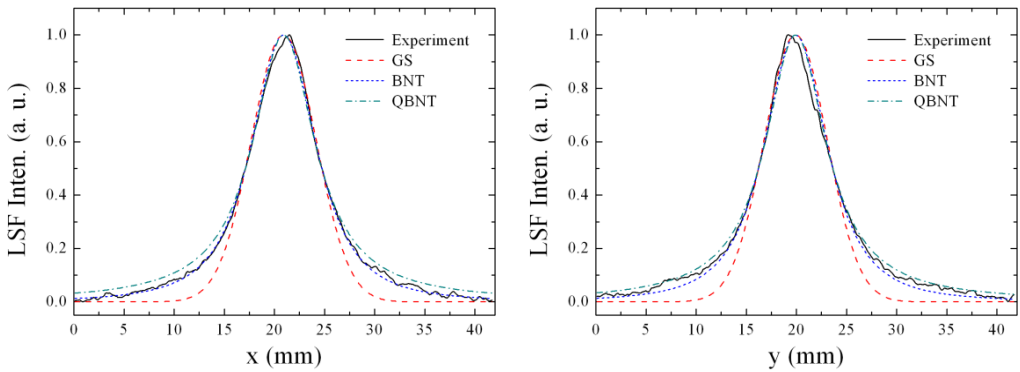

(a)

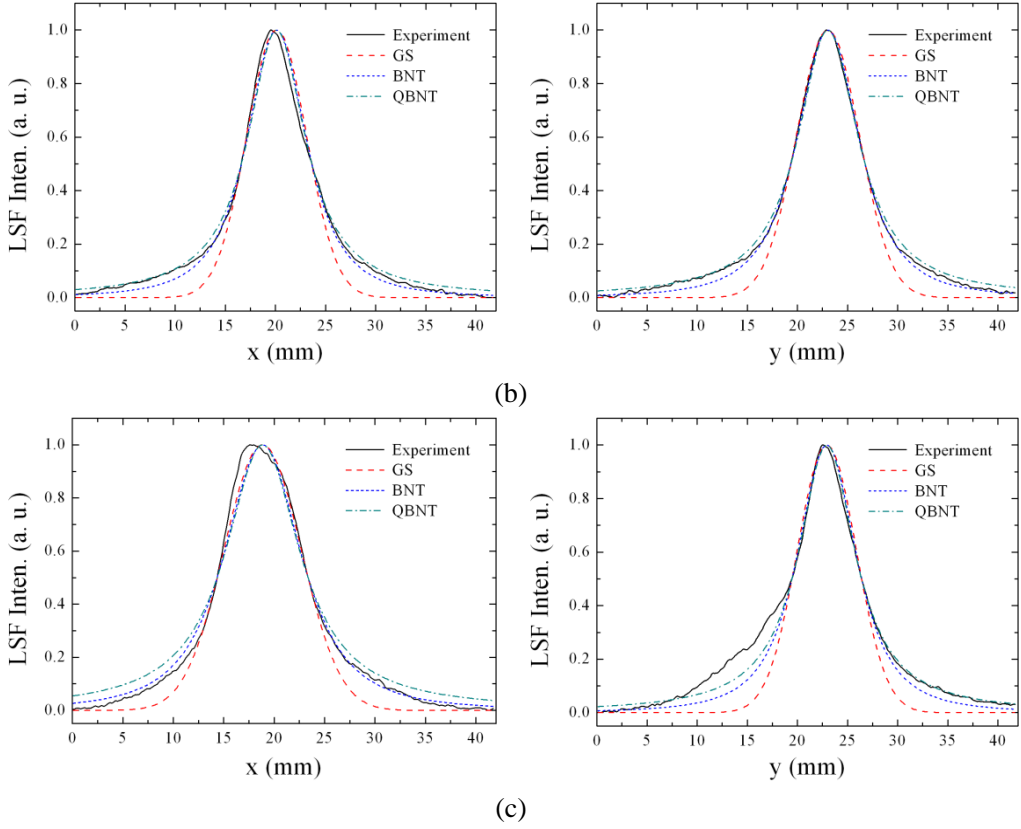

Fig. 3 (Color online) LSF curves of experimental images (No. 5295). (a) Pulse A; (b) Pulse B; (c) Pulse C.

Table 1 Experimental results of the spot size characterized by FWHM.

| No. | Pulse | E / MeV | I / kA | FWHM of PSF / mm | FWHM of LSF(x) / mm | FWHM of LSF(y) / mm | FWHM(LSF) / FWHM(PSF) |
|-----|-------|---------|--------|------------------|---------------------|---------------------|------------------------|
|      | A | 18.9 | 2.04 | 1.33 | 1.70 | 1.74 | 1.30 |
| 5293 | B | 18.9 | 1.98 | 1.35 | 1.53 | 1.78 | 1.22 |
|      | C | 18.8 | 2.07 | 1.45 | 1.77 | 2.34 | 1.41 |
|      | A | 19.0 | 2.03 | 1.26 | 1.77 | 1.74 | 1.39 |
| 5294 | B | 18.8 | 2.04 | 1.36 | 1.58 | 1.80 | 1.24 |
|      | C | 18.7 | 2.08 | 1.41 | 2.06 | 1.72 | 1.34 |
|      | A | 19.0 | 2.05 | 1.33 | 1.82 | 1.77 | 1.36 |
| 5295 | B | 19.0 | 2.08 | 1.41 | 1.68 | 1.77 | 1.22 |
|      | C | 18.9 | 2.09 | 1.50 | 2.16 | 1.66 | 1.28 |
|      | A | 19.0 | 2.06 | 1.33 | 1.78 | 1.76 | 1.33 |
| 5298 | B | 18.9 | 2.06 | 1.43 | 1.80 | 1.90 | 1.30 |
|      | C | 18.9 | 2.06 | 1.55 | 2.23 | 2.16 | 1.42 |
|      | A | 19.0 | 2.04 | 1.39 | 1.87 | 1.76 | 1.31 |
| 5299 | B | 19.0 | 2.06 | 1.45 | 1.91 | 1.82 | 1.29 |
|      | C | 18.9 | 2.07 | 1.62 | 2.16 | 1.89 | 1.25 |
|      | A | 19.0 | 2.05 | 1.35 | 1.76 | 1.85 | 1.34 |
| 5300 | B | 19.0 | 2.08 | 1.39 | 1.70 | 1.97 | 1.32 |
|      | C | 19.0 | 2.07 | 1.69 | 2.03 | 1.91 | 1.16 |

According to Eqs. (2) and (5), $D_L$ is calculated by finding the spatial frequency at half peak of the MTF. In order to deduct the screen blur, a 10-mm-thick tungsten plate is placed in contact with the screen and the penumbral image, which actually denotes the ESF of the screen blur, is recorded. Figure 4 shows the MTFs of the image LSF, the screen blur and the blur-deducted LSF for the triple-pulse spots in one measurement.

For each pulse, $D_L$ is calculated from the LSFs in the x and y directions, the Fourier transform of which correspond to the slices of the MTF in the spatial frequency space. Table 2 lists the experimental results of $D_L$ and their ratios to the PSF FWHM. It is seen that the difference between $D_L(x)$ and $D_L(y)$ is basically smaller than Pulses B and C, which agrees well with the result in the FWHM metric. The average ratio of $D_L$ to the PSF FWHM is 2.68 for Pulses A and B, 2.67 for Pulse C. This index reveals a spot distribution much closer to BNT, which is different from the indication of the LSF-to-PSF FWHM ratio.

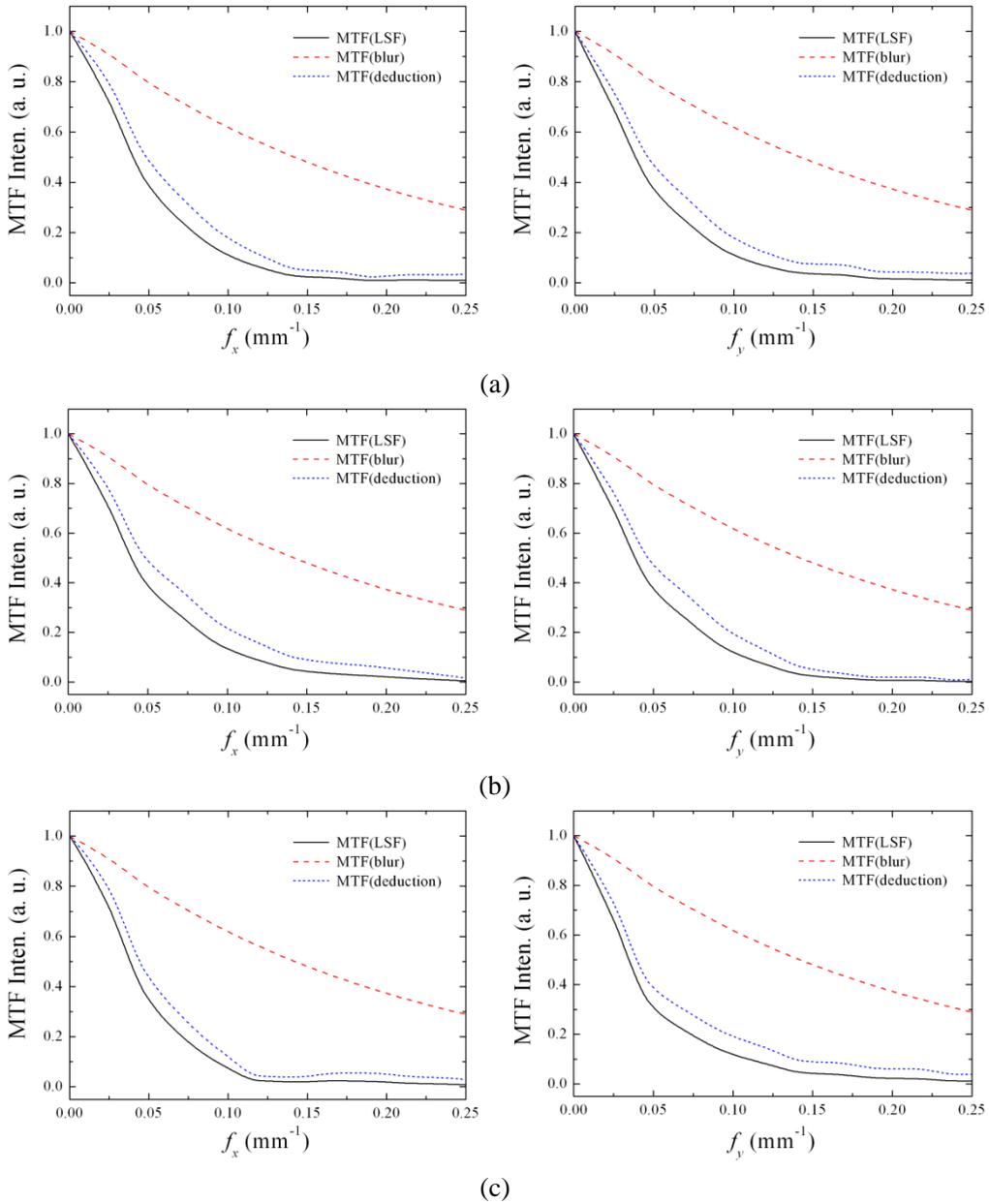

(a)

(b)

(c)

Fig. 4 (Color online) MTF curves of experimental images (No. 5295). (a) Pulse A; (b) Pulse B; (c) Pulse C.

Table 2 Experimental results of the spot size characterized by the LANL definition.

| No. | Pulse | $f_x$/mm$^{-1}$ | | $D_L(x)$/mm | $f_y$/mm$^{-1}$ | | $D_L(y)$/mm | $D_L$ / FWHM(PSF) |
| --- | --- | --- | --- | --- | --- | --- | --- | --- |
| | | with blur | without blur | | with blur | without blur | | |
| 5293 | A | 0.0409 | 0.0507 | 3.32 | 0.0402 | 0.0501 | 3.36 | 2.52 |
| | B | 0.0367 | 0.0444 | 3.80 | 0.0347 | 0.0409 | 4.12 | 2.92 |
| | C | 0.0387 | 0.0460 | 3.67 | 0.0322 | 0.0348 | 4.84 | 2.92 |
| 5294 | A | 0.0390 | 0.0470 | 3.59 | 0.0391 | 0.0475 | 3.54 | 2.83 |
| | B | 0.0387 | 0.0476 | 3.54 | 0.0384 | 0.0466 | 3.62 | 2.63 |
| | C | 0.0375 | 0.0439 | 3.84 | 0.0347 | 0.0380 | 4.43 | 2.94 |
| 5295 | A | 0.0399 | 0.0485 | 3.48 | 0.0379 | 0.0459 | 3.67 | 2.70 |
| | B | 0.0394 | 0.0486 | 3.47 | 0.0383 | 0.0466 | 3.62 | 2.51 |
| | C | 0.0400 | 0.0442 | 3.81 | 0.0341 | 0.0396 | 4.26 | 2.70 |
| 5298 | A | 0.0395 | 0.0473 | 3.56 | 0.0388 | 0.0469 | 3.59 | 2.69 |
| | B | 0.0360 | 0.0418 | 4.03 | 0.0372 | 0.0436 | 3.87 | 2.77 |
| | C | 0.0367 | 0.0402 | 4.19 | 0.0335 | 0.0365 | 4.62 | 2.84 |
| 5299 | A | 0.0381 | 0.0450 | 3.74 | 0.0384 | 0.0460 | 3.66 | 2.67 |
| | B | 0.0391 | 0.0466 | 3.62 | 0.0399 | 0.0486 | 3.47 | 2.45 |
| | C | 0.0400 | 0.0471 | 3.58 | 0.0385 | 0.0424 | 3.98 | 2.33 |
| 5300 | A | 0.0396 | 0.0473 | 3.56 | 0.0380 | 0.0457 | 3.68 | 2.69 |
| | B | 0.0370 | 0.0434 | 3.88 | 0.0368 | 0.0431 | 3.91 | 2.80 |
| | C | 0.0394 | 0.0464 | 3.63 | 0.0373 | 0.0408 | 4.13 | 2.29 |

The metrics of FWHM and LANL definition actually characterize the spot size either using a specific width of the spot intensity or choosing a select frequency in the MTF space, which may not show the overall picture. Here we also try to analyze the spot size and distribution in another way, calculating the edge radius and the weighted RMS radius ($R_{rms}$) of different ranges. The edge radius is given by the radius of an equal-area-disk, which is similar to the FWHM metric but corresponding to a certain PIS instead of half peak. The $R_{rms}$ takes into consideration the photon intensity and the distance to centroid of every position.

As shown in Fig. 5, the edge radius (relative to the PSF FWHM) of No. 5295 gradually turns from more like a QBNT distribution (10%PIS) to a BNT one (90%PIS). The $R_{rms}$ reveals a spot distribution between the BNT and QBNT for all PIS ratios, which, moreover, has much smaller discrepancy between the triple-pulse spots. Table 3 lists the experimental results of $R_{rms}$ at 90%, 50% and 10% PIS ratios. Theoretically, the $R_{rms}$ (relative to the PSF FWHM) at 90%PIS is 0.34 for KV, 0.52 for GS, 0.97 for BNT and 1.96 for QBNT. In the experiment, the average $R_{rms}$ at 90%PIS is obtained to be 1.51 mm for Pulse A, 1.63 mm for Pulse B and 1.59 mm for Pulse C, the ratio of which to the PSF FWHM is 1.14 for Pulse A, 1.17 for Pulse B and 1.04 for Pulse C.

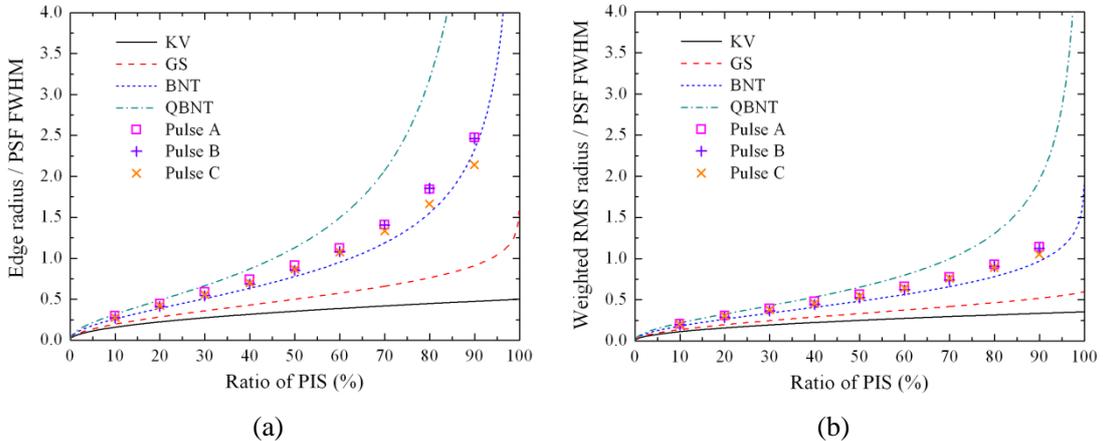

(a)                                          (b)

Fig. 5 (Color online) Radii of source spot with respect to the PIS boundary (No. 5295). (a) Edge radius; (b) weighted RMS radius.

Table 3 Weighted RMS radii within different PIS boundaries.

| No. | Pulse | $R_{rms}$ / mm | | | $R_{rms}$ / FWHM(PSF) | | |
|-----|-------|--------|--------|--------|--------|--------|--------|
| | | 90%PIS | 50%PIS | 10%PIS | 90%PIS | 50%PIS | 10%PIS |
| | A | 1.47 | 0.72 | 0.27 | 1.11 | 0.55 | 0.20 |
| 5293 | B | 1.69 | 0.77 | 0.27 | 1.25 | 0.57 | 0.20 |
| | C | 1.66 | 0.88 | 0.31 | 1.14 | 0.61 | 0.21 |
| | A | 1.51 | 0.74 | 0.27 | 1.20 | 0.59 | 0.21 |
| 5294 | B | 1.62 | 0.75 | 0.27 | 1.19 | 0.55 | 0.20 |
| | C | 1.59 | 0.82 | 0.29 | 1.14 | 0.59 | 0.20 |
| | A | 1.51 | 0.75 | 0.27 | 1.14 | 0.56 | 0.21 |
| 5295 | B | 1.58 | 0.74 | 0.27 | 1.12 | 0.52 | 0.19 |
| | C | 1.57 | 0.81 | 0.30 | 1.05 | 0.54 | 0.20 |
| | A | 1.51 | 0.75 | 0.27 | 1.13 | 0.56 | 0.20 |
| 5298 | B | 1.68 | 0.84 | 0.30 | 1.18 | 0.59 | 0.21 |
| | C | 1.72 | 0.93 | 0.33 | 1.11 | 0.60 | 0.21 |
| | A | 1.53 | 0.77 | 0.28 | 1.10 | 0.55 | 0.20 |
| 5299 | B | 1.58 | 0.77 | 0.29 | 1.09 | 0.53 | 0.20 |
| | C | 1.49 | 0.80 | 0.31 | 0.92 | 0.50 | 0.19 |
| | A | 1.54 | 0.77 | 0.28 | 1.14 | 0.57 | 0.21 |
| 5300 | B | 1.66 | 0.83 | 0.30 | 1.19 | 0.59 | 0.21 |
| | C | 1.51 | 0.81 | 0.32 | 0.89 | 0.48 | 0.19 |

The FWHM and $D_L$ are generally larger for Pulse C than the other two pulses, which are consistent with the $R_{rms}$ at a small PIS ratio. The reason is most probably due to a relative larger energy-spread of electron beam for Pulse C [16]. The energy-spread of the electron beam comes from imperfect square voltage waveforms supplied by the accelerating cavities and the asynchrony between them. The triple-pulser of the accelerator consists of three sets of square-voltage pulser, a confluent/blocking network, and a modular unit for triggering [17]. The reflection of the former voltage pulser(s) will add on the latter one(s). Therefore, the last electron beam pulse usually has the largest energy-spread.

5.  **Conclusion**

The triple-pulse X-ray source spot of the Dragon-II LIA is measured by means of the pinhole imaging method. The FWHM of the spot that is given by the diameter of an equal-area-disk and the spot size in the LANL definition are analyzed and compared with theoretical distributions. Experimental results show that Pulse C generally has a larger spot size and a greater asymmetry most probably because of a larger energy-spread of the electron beam and the worst homogeneity of the convertor target for the last pulse. The FWHM ratio of LSF to PSF indicates a QBNT distribution of the source spot while the ratio of $D_L$ to the PSF FWHM is much closer to the value of BNT. Moreover, the weighted RMS radius at different PIS ratios are calculated and analyzed for the pulses, which directly considers the contribution of every single photon within the boundary. And it reveals a spot distribution between the BNT and QBNT in a large scale of PIS ratio (from 10% to 90%PIS).